\documentclass[a4paper,10pt]{article}

\usepackage[utf8x]{inputenc}	
\usepackage[T1]{fontenc}

\usepackage{amsmath}
\usepackage{amsfonts}
\usepackage{latexsym}
\usepackage{amssymb}
\usepackage{bm}

\usepackage[pdftex]{graphicx} 

\usepackage{hyperref}

\title{Fifty years of Hubbard and Anderson lattice models: from magnetism to unconventional superconductivity - A brief
overview}
\author{%
J\'{o}zef Spa{\l}ek \\\vspace{6pt}\\
Email: ufspalek@if.uj.edu.pl \\\vspace{6pt}\\
Marian Smoluchowski Institute of Physics,\\Jagiellonian University, ul. Reymonta 4,\\
PL-30-059 Krak\'ow, Poland}

\begin{document}

\maketitle

\begin{abstract}
We briefly overview the importance  of Hubbard and Anderson-lattice models as applied to explanation of high-temperature
and heavy-fermion superconductivity. Application of the models during the last two decades provided an explanation of
the paired states in correlated fermion systems and thus extended essentially their earlier usage to the description of
itinerant magnetism, fluctuating valence, and the metal-insulator transition. In second part, we also present some of
the new results concerning the unconventional superconductivity and obtained very recently in our group. A comparison
with experiment is also discussed, but the main emphasis is put on  rationalization of the superconducting properties of
those materials within the real-space pairing mechanism based on either kinetic exchange and/or Kondo-type interaction
combined with the electron correlation effects.
\end{abstract}

\section{Introduction: Hubbard- and Anderson-lattice models}

The main purpose of this paper is to emphasize the conceptual development of the models specified in the title starting
from the description of itinerant magnetism and correlated electron states in the normal state, into a unified framework
encompassing also unconventional superconducting paired states and the coexistent magnetic-superconducting phases. We
start with a general historic note on the role of second quantization.

Principal development of the so-called quantum many body physics was strongly influenced by a wide application of the
second-quantization methods. The pioneering seems to be the work of Holstein and Primakoff  \cite{b1} on the magnon
(spin-wave) excitations in the (broken symmetry) Heisenberg ferromagnetic state, later extended to the two-sublattice
antiferromagnets \cite{b2}. The success of the theoretical approach was related to the fact that these collective
excitations of the system of interacting localized magnetic moments can be reduced to the first approximation to a
quantum system of coupled harmonic oscillators, leading thus to elementary excitations for which the residual coupling
between them produces their finite but small lifetime, but this is an effect of higher order, at least deep inside of
the broken-symmetry state (i.e., at low temperatures, $T\ll T_{c}$). This formalism has its origin in the quantum-field
theoretical representation of the spin via bosons \cite{b3}. The other pioneering 
development was the Bogoliubov microscopic approach to the condensation and excitations of a weakly interacting Bose gas
of material particles \cite{b4}. In other words, the interacting system was represented by a system of weakly
interacting quasiparticles, "quasi" meaning their appearance in the condensed state (ferromagnetic, superfluid, etc.),
as well as the fact that they have a finite lifetime for temperature $T>0$. One should note that quasiparticles appear
as true particles when we consider e.g., scattering of neutrons on spin waves or studying sound excitations in a
superfluid. Thus, the second-quantization scheme expresses among others things, return to the particle language in the
correct formulation by quantizing the classical- or matter-wave dynamics (hence the phrase \emph{second quantization}).

The corresponding development for fermions, which compose most of our materials world, culminated in the development of
the Bardeen, Cooper, and Schieffer (BCS) theory of superconductivity \cite{b5,b6}. This approach has had as a
predecessor the works of Fr\"{o}hlich \cite{b7} and that of Bardeen and Pines \cite{b8}. Those works concentrated on the
origin of attractive interaction leading to the superconducting condensation in an electron gas, though the works on an
individual particle "dressed" by a cloud of virtual phonons (the polaron) should also be noted \cite{b9}, as it helped
to formulate the concept of fermionic quasiparticle in the normal Fermi liquid \cite{b10,b11}.

All those concepts, deeply rooted in the second-quantization language, were based on either the harmonic-oscillator
(bosons) or the ideal electron-gas (fermions) many-particle states and associated with them statistics of counting their
occupations (the Bose-Einstein and the Fermi-Dirac distributions, respectively). A completely fresh start, in a specific
solid-state context, is associated with the two works of Anderson: the first on the kinetic exchange for the
antiferromagnetic  (Mott) insulators \cite{b12,b13} (to which the first formulation, albeit implicitly, of the Hubbard
model can be traced) and the second, on magnetic impurity in a normal metal \cite{b14}. Although those works were the
first modeling electrons in correlated solid-state materials, starting from an atomic representation of solids and the
realistic Mott insulating state \cite{b15}, the pioneering character of the works on correlated systems is associated
nowadays with the papers of Hubbard \cite{b16,b17,b18}. Such situation is probably 
due to the three circumstances. First, the Hubbard devised a simple model of itinerant electrons, with the help of which
the phenomenological Stoner model of magnetic state of fermions \cite{b16} could be rationalized. Second, most
importantly, the localization-delocalization-transition (insulator-metal transition) concept of Mott, based on
electron-gas picture \cite{b17} has been put on a firm basis of narrow-band systems, regarded as complementary metallic
systems to the corresponding electron gases. Third, the conceptual ingenuity of the Anderson-impurity model has come to
light only after the seminal paper of Schrieffer and Wolff \cite{b18} on its canonical transformation to the so-called
Kondo model \cite{b19}, and first of all, after the extension of this model to the periodic (Anderson- or Kondo-lattice)
systems \cite{b20,b21,b22,b23}, where the latter model plays a dominant role of interpreting the data for heavy-fermion
and related magnetic systems to this day. To recapitulate, the Hubbard- and the 
Anderson-lattice models are regarded as universal models in condensed matter physics \cite{b16,b24,b25}, including their
orbitally degenerate versions (Refs. \cite{b26} and \cite{b27,b28}, respectively).

Those two models (together with the subsidiary Kondo-lattice and the s-d-type models) are examples of the parameterized
models. By that we mean that the model parameters are expressed via single-particle (band) structure in the
thigh-binding approximation (via the hopping and the hybridization parameter(s), $t_{ij}$ and $V_{ij}$, respectively),
as well as by the short-range (mostly intraatomic) part of the interparticle interaction, i.e., with the help of the
following interaction parameters: the Hubbard $U$ (intraatomic intraorbital Coulomb), $U'$ or $K$ (the intraatomic
interorbital or the intersite in one-orbital case, respectively), and $J$ (the ferromagnetic intersite or the intrasite
interorbital interaction, the Hund's rule exchange). Those parameters contain integrals over the single-particle wave
functions of the states among which we include the interaction, whereas the remaining operator part (in the
second-quantization scheme) describes the interparticle correlations. These interaction parameters 
are usually  regarded as independent of detailed dynamics in the second-quantization language, which is not always the
case.

\subsection{A brief methodological note}

The single-particle wave functions (Wannier or Bloch functions), selected to define the field operator, as well as to
define the model, are usually regarded as independent of the degree of correlation between quasiparticles. This is not
necessarily the case, particularly in the regime, where the nature of single-particle states changes, e.g., at the
metal-insulator transition (the Mott-Hubbard localization). This is because if the single-particle bases selected to
define a model were complete, their choice would be absolutely arbitrary. However, in the cases of either the Hubbard or
the periodic Anderson models this is not so. Therefore, the basis should be, in our view, optimized in accord with the
situation at hand. In the series of papers \cite{b29,b30,b31,b32,b33} we have addressed this question in detail in some
model situations. Such reformulation of the model allowed, among others, to determine \cite{b32} a relation between the
Mott and the Hubbard criteria for metal-insulator transformation.

\subsection{Structure of the paper}

After the introduction (cf. above), we address in Sec. 2 the question of the t-J model emergence, originally from the
Hubbard model and introduce real-space pairing in general terms. Next, in Sec. 3 we discuss physical results concerning
the superconducting state within t-J model. In Sec. 4 we discuss the principal features of the pairing in te
Anderson-Kondo lattice model. Sec. 5 contains concluding remarks and an outlook. Figures have been assembled into panels
and thought to substitute a detailed formal discussion concerning the details of a formal solution. No detailed account
of the relevant references is undertaken.

\section{From Hubbard model to t-J model: from magnetism to superconductivity}

\subsection{General remarks}

As has been already said, the Hubbard model has been used in the first two decades (1963-1985) to rationalize the
itinerant magnetism of electrons in narrow correlated bands. By correlated systems we understand the systems for which
the so-called bare bandwidth $W$ (or equivalently, the Fermi energy $\mu$ measured from the bottom of the narrow band)
is comparable or even substantially smaller than the principal parameter - magnitude $U$ of the Hubbard interaction
between the particles with opposite spins and located on the same Wannier orbital. Additionally, the model has been used
extensively to discus the metal-insulator transition taking place for a half-filled band i.e., for one particle per
site, $n=1$, as a function of the interaction-to-bandwidth ratio $U/W$.

A separate question concerns the kinetic-exchange interaction derivation from the Hubbard  model. Namely, how to
generalize it derived originally for the case of Mott insulator \cite{b13,b34} to the case of strongly correlated metal,
i.e., for both $U/W \gg 1$ and for a arbitrary band filling (say, $n \leq 1$). In this respect, our original description
\cite{b35,b36,b37}  filled the gap by defining also explicitly the strongly correlated state which represents the
starting point to high-temperature superconductivity within the so-called t-J model \cite{b38,b39,b40} and associated
with it  concepts of real-space pairing and of the so-called \emph{renormalized mean-field theory} (RMFT), which we are
going to discuss next. But first, we write few general remarks about the Hubbard model.

\subsection{The Hubbard model as such}

As already said, the new feature of Hubbard's original reasoning \cite{b16,b17,b18} was to start with the description of
interacting electrons (fermions) in terms of atomic or Wannier states on a lattice rather than from an interacting
electron gas. In this manner, try to describe the metals close to the Mott (localized) state. Explicitly, the single
narrow-band Hubbard Hamiltonian reads
\begin{equation}
\mathcal{H} = \sum_{\langle ij\rangle\sigma}\!^{'}t_{ij}\,\hat{a}_{i\sigma}^{\dagger}\,\hat{a}_{j\sigma}+U\sum_{i}
n_{i\uparrow}\, n_{i\downarrow}\equiv
 \sum_{\pmb{k}\sigma}\epsilon_{\pmb{k}}\, n_{\pmb{k}\sigma}+U\sum_{i}n_{i\uparrow}\,n_{i\downarrow},
 \label{r1}
\end{equation}
where
\begin{equation}
t_{ij}\equiv\langle w_{i}|H_{1}|w_{j}\rangle\equiv \int d^{3}r\,w_{i}^{*}(\pmb{r})\, \mathcal{H}_{1}(\pmb{r})\,
w_{j}(\pmb{r})
\end{equation}
is the hoping integral ($i\neq j$) expressed in terms of the single-particle Hamiltonian $\mathcal{H}_{1}(\pmb{r})$ and
a set of single-particle Wannier orbitals $\{w_{i}(\pmb{r})\}$ for this single band, that are composed of atomic states
$\left\{ \Phi_{i}(\pmb{r})\right\}$ in the following manner
\begin{equation}
w_{i}(\pmb{r}) = \sum_{j} c_{ij}\,\Phi_{i}(\pmb{r}),
\end{equation}
with $\sum_{j}|c_{ij}|^{2}=1$. The interaction (Hubbard) parameter $U$ is of the form
\begin{equation}
U\equiv\left\langle w_{i}\,w_{i}\left| V
\right|w_{i}\,w_{i}\right\rangle\equiv
\int d^{3}r\,d^{3}r'\left|w_{i}(\pmb{r})\right|^{2}V(\pmb{r}-\pmb{r}') \left|w_{i}(\pmb{r}')\right|^{2},
\end{equation}
where $V(\pmb{r}-\pmb{r}')$  represents two-particle (usually repulsive) interaction in the coordinate representation.
$U$ plays the crucial role in determining the system properties depending on its relative value with respect to the bare
bandwidth $W\equiv2\left|\sum_{j(i)}t_{ij}\right|$ of the starting band states with the dispersion relation
$\epsilon_{\pmb{k}}$. Namely: (i) for $U\ll W$ we have \emph{a metallic limit} with single-particle narrow-band states
in the thigh-binding approximation, with the itinerant-electron magnetic and paramagnetic states being represented in
the Hartree-Fock approximation (among them, the Stoner criterion and the dynamic susceptibility in RPA approximation);
(ii) for $U\gg W$ we have \emph{a complementary strong-correlation limit}, where the kinetic exchange interaction
describes properly the antiferromagnetic state of the Mott insulator for $n=1$ and the state of strongly correlated
metal for $n \neq 1$; and
(iii) for $U\approx W$ we have for a half-filled band a transition from an moderately correlated metallic state to
either Mott insulator (for $n=1$) or to a strongly correlated metal. The regime (ii) and (iii) are the most interesting,
but most difficult to tackle within a single formal scheme.

The truncation of the full interparticle interaction in (\ref{r1}) to the intraatomic (intraorbital) part means that we
can describe the single-orbital systems with lattice parameter $a\ll a_{B}$, where $a_{B}$ is the characteristic length
(the effective Bohr radius of the single-particle atomic states in the medium usually). In that situation
$|c_{ij}|\ll|c_{ii}|$ for $i\neq j$. Under this condition $t_{ij}$ can be usually limited to that between the nearest
$\langle ij \rangle$ and the next-nearest neighbors. In effect, the two parameters $t_{\langle ij\rangle}$ and $U$
define \emph{the tight-binding approach} (\ref{r1}) in a general sense for \emph{narrow-band systems}.

\subsection{Hubbard model: weak to moderate correlations}

As we are interested mainly in real-space pairing and associated with it superconductivity, let us make here few remarks
summarizing very briefly the effort in the first two decades of the model studies. It has been noted already by Hubbard
in its original paper in 1963 \cite{b16} that fulfilling the Stoner criterion for the onset of ferromagnetism is not
easy. This is because one can say intuitively that within the orbitally nondegenerate Hubbard model there is no obvious
ferromagnetic exchange contribution (such a is provided by the Hund's rule in the degenerate-band case), which would
stabilize a homogeneous spin-polarized state. The usual argument quoted in this situation is that the parallel-spin
state is stabilized by electronic correlations, since the repulsive interaction suppresses double occupancies with the
opposite spins and hence supports implicitly the intersite parallel-spin configuration, as then the
repulsive-interaction energy $\sim U\langle \hat{n}_{i\uparrow}\,\hat{n}_{i\downarrow}\
rangle$ is reduced automatically. However, this is not quite true as, at least near the half-filling, the
antiferromagnetic (Slater) state is stable in the case of bipartite lattice, whereas the itinerant ferromagnetic state
becomes stable only for a substantially higher $U/W$, far beyond the value, where the Hartree-Fock approximation would
be realistic \cite{b41,b42}.

The other characteristic feature introduced Hubbard \cite{b17}, appeared in the subsequent papers in the regime of
$U\simeq W$, i.e., in the regime of the metal-insulator transition. Here, the original solution of Hubbard \cite{b17}
provided the concept of a spontaneous splitting into two (Hubbard) subbands of a single-particle band even in the
paramagnetic state, which produced microscopically for the first time an insulating state induced by the correlations.
This result contrasts with the original Wilson (1932) classification of solids, within which a band system with an odd
(here one per site) number of electrons should be always a metal. The last conclusion is however in direct contradiction
with properties of e.g., CoO (with $3d^{7 } 2p^{6}$ valence-band configuration) which is one of very good insulators,
albeit an antiferromagnet. Parenthetically, this insulating system cannot be regarded as a split-band Slater
antiferromagnet, as it remains a good insulator (or a wide-gap semiconductor) well above 
its N\'{e}el temperature $T_{N}\simeq 290$ K.

In recent years the band-theoretical approaches with inclusion of correlations: LDA+U \cite{b43,b44} and LDA+DMFT
\cite{b45,b46} were widely used and incorporate the correlations into the local-density approximation (LDA) scheme.
Nonetheless, they seem not to provide as yet systematic answers, particularly for low-dimensional systems such as e.g.,
La$_{2}$CuO$_{4}$. In effect, the parametrized models such as Hubbard t-J or Anderson-Kondo lattice models are still
good alternatives, particularly when discussing high temperature superconducting or heavy-fermion properties, i.e.,
phenomena at low energies (low temperatures).

Apart from the methods used above, we would like to quote the variational method of Gutzwiller wave function (GWF)
\cite{b47,b48}, which represented a contemporary to Hubbard (1963-65)  and an independent alternative formulation, also
with respect to the Hubbard Hamiltonian introduction. This method has been analyzed extensively starting from the work
of Brinkman and Rice in 1970 \cite{b49}, who showed that a simpler Gutzwiller approximation (GA) produces a divergent
static magnetic susceptibility at the metal-insulator transition (MIT). The divergence displayed in Fig. 1 for example
of constant density of states, with the critical value for the metal - insulator transition $U_{C}=2W$, what proves that
MIT is realized as a continuous quantum phase transformation. Although, the Gutzwiller approximation is exact only for a
 lattice of infinite dimensions \cite{b50,b51}, it provides to this day a qualitative border point dividing the
narrow-band systems into weakly or moderately correlated systems from one side 
and those being in the strong correlation regime from the other.

Before considering the strongly correlated systems, we should note one more very important qualitative feature of the
Gutzwiller approximation, which survived to these days. Namely, it allowed to introduce a self-consistent procedure of
introducing the concept of quasiparticle (the so-called \emph{statistical quasiparticle}) in the regime $U\gtrsim W$. In
this regime, the Landau theory of Fermi liquids in its standard version is inapplicable. In essence, the renormalized
quasiparticle energies have the form $E_{\pmb{k}\sigma}=q_{\sigma}\,\epsilon_{\pmb{k}}$, where the band narrowing factor
$q_{\sigma}$ is related to the effective mass enhancement, i.e., $m^{*}_{\sigma}/m_{B}=1/q_{\sigma}$, where $m_{B}$ is
the bare band mass. Explicitly, we have
\begin{equation}
q_{\sigma}=\frac{\sqrt{n_{\sigma}-d^{2}}\,\sqrt{1-n+d^{2}}+d\sqrt{n_{\bar{\sigma}}-d^{2}}}
{n_{\sigma}(1-n_{\sigma})},
\label{r5}
\end{equation}
where $n_{\sigma}\equiv \langle\hat{n}_{i\sigma} \rangle$ and
$d^{2}\equiv \langle\hat{n}_{i\uparrow}\, \hat{n}_{i\downarrow} \rangle$ is the probability of having double occupancy
which is determined by minimizing the ground state energy of the system
\begin{equation}
E_{G}/N=\frac{1}{N} \sum_{\pmb{k}\sigma} E_{\pmb{k}} \left\langle\hat{n}_{\pmb{k}\sigma} \right\rangle+ Ud^{2},
\end{equation}
where at $T=0$  $\langle\hat{n}_{\pmb{k}\sigma} \rangle = \Theta(\mu-E_{\pmb{k}\sigma})$ and $\mu$ is the chemical
potential. The parameter $d^{2}$ reaches zero at $U=U_{C} \equiv 8|\bar{\epsilon}|$, where
\begin{equation}
\bar{\epsilon}=\frac{1}{N} \sum_{\pmb{k}\sigma} \epsilon_{\pmb{k}}\left\langle\hat{n}^{0}_{\pmb{k}\sigma} \right\rangle,
\label{r7}
\end{equation}
and $\left\langle\hat{n}^{0}_{\pmb{k}\sigma} \right\rangle \equiv  \Theta(\mu^{0}-\epsilon_{\pmb{k}\sigma})$; with
$\mu^{0}$ being the chemical potential for bare electrons and $\langle\ldots\rangle$ the statistical distribution and
the Fermi energy for bare electrons. In Figs. 2 and 3 we plot the mass enhancement factors for the spin-polarized and
paramagnetic states, respectively. Note that the quasiparticle mass is explicitly spin-direction dependent. The
quasiparticles in the spin-minority subband become heavier with the increasing polarization, since they scatter on a
larger number of (spin-majority) quasiparticles. For the majority-spin carriers the opposite is true.

This picture of self-consistently determined quasiparticle characteristics leads to a number of unique physical
properties among them a strong metamagnetism \cite{b52} and a nonstandard temperature dependence of the metal-insulator
transition \cite{b53,b54}. What is more important, with the introduction of the spin-resolved mass differentiation
(i.e., the increasing polarization) we reach the limit of distinguishable quasiparticles, for $m^{*}_{\uparrow}\neq
m^{*}_{\downarrow}$, out of indistinguishable particles ($m^{*}_{\uparrow}=m^{*}_{\downarrow}= m_{B} $) when the
starting state at zero applied magnetic field is paramagnetic \cite{b55}. These properties represent a point of
departure to the analysis in the strong-correlation limit $U\gg W$. Note that GA in this standard form provides only
properties up to maximal value $U=U_{C}\sim W$ (cf. Figs. 1-3). Similar properties can be obtained within the periodic
Anderson model in the strong correlation limit \ \cite{b56}. Analogical conclusions can be obtained 
also for orbitally degenerate systems  \cite{b57}.

In relation to what has been said above, two methodological remarks are in place. First,
the Gutzwiller approximation has been subsequently reformulated in the quantum-field-theoretical languages as the
so-called slave boson approach  \cite{b58,b59,b60}. Within this approach, GA is regarded as a saddle-point
approximation. While relying on the GA results in its initial stage, the slave-boson approach removes one of the
principal inaccuracies of GA. Namely, within GA the self-consistent procedure of calculating the averages (particularly,
in this spin-polarized state) provides results which differ from those obtained from an appropriate variational
procedure. This deficiency of the method has been corrected with the introduction of \emph{statistically consistent
Gutzwiller approximation} (SGA), which not only brings into agreement the original results and those of the slave-boson
approach in the saddle point approximation, but also avoids introducing the slave (ghost) boson fields which are
introduced ad hoc in the latter formulation  \cite{b61}. In this manner, a consistent mean-field treatment 
has been formulated, which will be discussed in detail in the context of unconventional superconductivity in the
strong-correlation limit which as discussed next.

The reformulation of the Gutzwiller approach in the quasiparticle language \cite{b58,b59,b60} has one additional
advantage. Namely, it is applicable to the $T>0$ situation.

Second, as the GA approximation has, strictly speaking, precise meaning for high-dimension ($d\rightarrow \infty$)
limit, the Brinkman-Rice analysis is often regarded as qualitative at best, i.e., setting the division into regimes of
$U<U_{c}$ and $U>U_{c}$ as regimes with qualitatively different physics in each of them. There have been various
analytic and numerical trials to allow for an interpolation between those two limits. Apart from extensive an Quantum
Monte Carlo analysis, a generalization of the Gutzwiller wave function to include double-holon correlations was proposed
\cite{b32a}. This approach provides better ground state energy and characteristic while preserving the principal physics
of GA. However, it leads to the first-order Mott transition, in agreement with our recent SGA analysis \cite{b33}.

\subsection{Strong correlation limit: t-J model}

As said above, the starting half-filled-band metal transforms for $U\gtrsim W$ into the Mott (or more precisely,
Mott-Hubbard) insulator, which is essentially an antiferromagnet with localized moments. Anderson \cite{b62} was the
first to point out that the Mott insulating state of e.g., La$_{2}$CuO$_{4}$ with $3d^{9}$ (spin $S=1/2$) configuration
of Cu$^{2+}$ ions can be regarded as the parent material for the high temperature superconductor
La$_{2-x}$Sr$_{x}$CuO$_{4}$ (LSCO) which appears for the doping $x \gtrsim 0.05$, with the deficient electrons
effectively producing Cu$^{2+x}$ configurations of itinerant holes in this doped Mott insulator. The effective
Hamiltonian known nowadays under the acronym of "t-J model" was derived originally by us \cite{b35,b36,b37} for an
arbitrary band filing $n$ (hole doping $x \equiv 1-n$) and has the form
\begin{eqnarray}
\widetilde{\mathcal{H}} = \sum_{ij\sigma}\!^{'}t_{ij}\,\hat{b}_{i\sigma}^{\dagger}\,\hat{b}_{j\sigma}+\sum_{ij}\!^{'}
\frac{2t_{ij}^{2}}{U}\left(\widehat{\pmb{S}}_{i}\cdot \widehat{\pmb{S}}_{j}-\frac{c_{1}}{4}\,\widehat{\nu}_{i}\,
\widehat{\nu}_{i} \right) \nonumber \\
+ \sum_{ijk\sigma}\!^{''}\,\frac{t_{ij}\,t_{jk}}{U}\,c_{2}
\left(\hat{b}_{i\sigma}^{\dagger}\, \widehat{\nu}_{j\bar{\sigma}}\,\hat{b}_{k\sigma}- \hat{b}_{i\sigma}^{\dagger}\,
\widehat{S}_{j}^{\bar{\sigma}}\,\hat{b}_{k\bar{\sigma}}
\right),
\label{r8}
\end{eqnarray}
where the single-primed summation means that $i\neq j$, the double-primed means that $i\neq j\neq k\neq i$. Also, 
$\hat{b}_{i\sigma}^{\dagger} \equiv \hat{a}_{i\sigma}^{\dagger}(1-\widehat{n}_{i\bar{\sigma}})$,
$\hat{\nu}_{i\sigma}\equiv \hat{b}^{\dagger}_{i\sigma}\,\hat{b}_{i\sigma}=
\hat{n}_{i\sigma}(1-\widehat{n}_{i\bar{\sigma}})$ are the projected fermionic operators, $\widehat{\pmb{S}}_{i} \equiv
(\widehat{S}_{i}^{z}, \widehat{S}_{i}^{\sigma})$ is the spin operator in the fermion representation. The first term
represents a restricted hopping, with the double occupancies projected out, the remaining two terms represent
respectively the second-order processes and contain virtual-hopping processes to the double occupancies configuration
(for the didactical exposition see e.g., \cite{b63,b64}). The extra parameters  $c_{1}, c_{2}=0,1$ are introduced to
model the system with and without of the corresponding terms respectively \cite{b65}.

At the times, this Hamiltonian was used to describe mainly magnetic properties (mainly ferromagnetism)  at or near the
Mott insulating limit \cite{b66,b67,b68}. In other words, only the moving correlated spins have been seen in this
unusual situation with the double-site occupancies being ruled out . As those moving spins avoid each other
($\langle\widehat{n}_{i\uparrow}\,\widehat{n}_{i\downarrow}\rangle \equiv 0$), the intersite antiferromagnetic
interaction (the second term) produces strong pair spin-singlet correlations, that can produce either a long-range
antiferromagnetism, or a liquid of pair spin singlets entangled with the hole hopping. This last state is sometimes
termed \emph{a resonating-valence-bond (RVB) state} \cite{b69,b70}. However, since in the correlated state the
renormalized hoping magnitude is $t_{ij} \rightarrow t_{ij}x$ and the corresponding kinetic-exchange integral is
$J_{ij}\equiv 2t_{ij}^{2}/U \rightarrow J_{ij}(1-x)^{2}$, for $x\lesssim~0.1$ the exchange term becomes predominant and
the hopping strongly suppressed, what produces an insulating antiferromagnetic state as $x\rightarrow 0$. Such was the
state of affairs until 1986.

The invention of real space pairing \cite{b62,b38,b39} has introduced the completely new aspects to the problem. First,
one can introduce explicit the real-space spin-singlet pairing operators in a rigorous manner in the form \cite{b39}
\begin{align}
\left\{
\begin{array}{l}
B_{ij}^{\dagger}\equiv \frac{1}{2}\left(\hat{b}_{i\uparrow}^{\dagger}\:\hat{b}_{j\downarrow}^{\dagger}-
\hat{b}_{i\downarrow}^{\dagger}\:\hat{b}_{j\uparrow}^{\dagger}\right), \\
B_{ij}\equiv \frac{1}{2}\left(\hat{b}_{i\uparrow}\:\hat{b}_{j\downarrow}-
\hat{b}_{i\downarrow}\:\hat{b}_{j\uparrow}\right)
\end{array}
\right.
\end{align}
and cast the effective Hamiltonian (\ref{r8}) in the following closed form \cite{b39}
 \begin{align}
\widetilde{\mathcal{H}}= \sum_{ij\sigma}\!^{'}  t_{ij}\:\hat{b}_{i\sigma}^{\dagger}\:\hat{b}_{j\sigma} -
\sum_{ijk}\!^{'}\,  \frac{4 t_{ij}\:t_{jk}}{U} B_{ij}^{\dagger}\: B_{kj}.
\label{r10}
\end{align}

From this expression one can see explicitly that a nonzero number of local singlet pairs $\langle B_{ij}^{\dagger}\,
B_{ij}\rangle$ diminishes the system energy and additionally, in the Hartree-Fock-like factorization of the Bogoliubov
type: $\langle B_{ij}^{\dagger}\rangle\, \langle B_{ij}\rangle$, the quantity $\langle B_{ij}^{\dagger}\rangle$ is the
real-space correspondant of the order parameter in the Barden-Cooper-Schrieffer (BCS) theory, $\Delta_{\pmb{k}}\equiv
\langle \hat{a}_{\pmb{k}\uparrow}^{\dagger}\, \hat{a}_{-\pmb{k}\downarrow}^{\dagger}  \rangle$. However, there is one
principal difference: in the present situation we have $\langle B_{ii}^{\dagger}\rangle \equiv 0$ rigorously due to the
Gutzwiller projection and therefore, the gap parameter cannot have a scalar ($\pmb{k}$-independent) $s$-wave form.
Instead, in the simplest situation it must have either the extended $s$-wave or the $d$-wave form \cite{b71}. These
forms are the simplest representations of the intersite character of the pairing 
amplitude $\langle B_{ij}^{\dagger}\rangle$.

The second factor was the discussion of electronic states for CuO$_{2}$ plane in LSCO a subsequent reintroduction
\cite{b62,b40} of the t-J model with $J_{ij}$ not limited to the asymptotic expression for $|t_{ij}|\ll U$ appearing as
(\ref{r8}). In effect, that value of the ratio $J/|t|$ within such redefined  effective t-J model is usually taken for
the nearest neighbors as $J/|t|=0.3$. This ratio value is assumed in the following, when illustrating the theoretical
results with a detailed numerical analysis.

A methodological note is in place here. The representations (\ref{r8}) and (\ref{r10}) are simply equivalent. Hence the
motion of the local singlets in the paired state, as provided by the second term in (\ref{r10}) and the correlated
hoping of holes must cooperate with each other in forming such a condensed state of moving pairs. Such cooperation of
the two factors has been proposed recently \cite{b73}.

\section{Description of the high-$T_{C}$ superconducting state}

\subsection{Statistically consistent mean-field approach (SGA)}

The standard approach to the description of superconducting state starting from (\ref{r10}) is the so-called
\emph{renormalized mean field theory} (RMFT) \cite{b72,b73,b74}. However, as we have noticed \cite{b61,b65,b76,b77} the
additional constraints must be reintroduced so that the averages appearing in the effective single-particle Hamiltonian
and calculated self-consistently coincide with those determined by an alternative variational approach. This is the
basic principle which must be obeyed by any consistent approach from the statistical physics point of view. In result,
the effective single-particle Hamiltonian for $c_{1}$ and $c_{2}=0$ reads:
\begin{align}
\mathcal{H}=&\sum_{ij\sigma}\left( t_{ij}\,g_{ij}^{t}\,\hat{a}_{i\sigma}^{\dagger}\,\hat{a}_{j\sigma}+\text{H.c.}\right)
-\mu\sum_{i\sigma}\hat{a}_{i\sigma}^{\dagger}\,\hat{a}_{i\sigma} \nonumber\\
-&\frac{3}{4}\sum_{ij\sigma} J_{ij}\,g_{ij}^{J}
\left(\chi_{ij}\,\hat{a}_{i\sigma}^{\dagger}\,\hat{a}_{j\sigma}+\text{H.c.}-|\chi_{ij}|^{2}\right)\nonumber\\
-&\frac{3}{4}\sum_{ij\sigma} J_{ij}\,g_{ij}^{J}
\left(\Delta_{ij}\,\hat{a}_{i\sigma}^{\dagger}\,\hat{a}_{j\bar{\sigma}}^{\dagger}+H.c.-|\Delta_{ij}|^{2}\right),
\label{r11}
\end{align}
where  $\chi_{ij}\equiv \langle\hat{a}_{i\sigma}^{\dagger}\,\hat{a}_{j\sigma}  \rangle$ and $\Delta_{ij}\equiv \langle
\hat{a}_{i\bar{\sigma}}\,\hat{a}_{i\sigma} \rangle $ are respectively the hopping amplitude and the pairing other
parameter. The renormalization factors $g_{ij}^{t}$ and $g_{ij}^{J}$ result from the Gutzwiller approximation.
Additionally, when the statistical-consistency conditions \cite{b76,b77} are included, the Hamiltonian can be written in
the form
\begin{align}
\mathcal{H}_{\lambda}&= W- \sum_{\langle ij\rangle\sigma} \widetilde{\eta}_{ij\sigma}
\left[ \left(\hat{a}_{i\sigma}^{\dagger}\,\hat{a}_{j\sigma}-\chi_{ij\sigma} \right)+\text{H.c.}\right]\nonumber\\
&- \sum_{\langle ij\rangle} \widetilde{\gamma}_{ij}
\left[ \left(B_{ij}-\Delta_{ij} \right)+\text{H.c.}\right]
-\sum_{i\sigma}\widetilde{\lambda}_{ni}\left(\widehat{n}_{i\sigma}-n_{i\sigma}\right),
\label{r12}
\end{align}
where $W\equiv \langle\mathcal{H} \rangle$ means the mean-field expectation value of (\ref{r11}) and the last three
terms represent the Lagrange constraints with the corresponding multipliers $\widetilde{\eta}, \widetilde{\gamma}$, and
$\widetilde{\lambda}$. In the case of spatially homogeneous state the solution reduces to solving a system of 6
algebraic (integral) equations. The results obtained in this manner have been displayed  in the panel composing Figs.
4b-d \cite{b65,b77}. For the sake of completeness, in Fig. 4a we present a schematic phase diagram obtained
experimentally and encompassing the doping regime $0.055 \lesssim x \lesssim 0.35$, where the superconducting state is
stable.

Few features of the whole approach should be noted. First, the approach contains the correlated gap parameter
$\Delta_{c}\equiv\langle B_{ij} \rangle\equiv g_{t}\,\Delta_{ij}$, the correlated hopping amplitude $\langle
(g^{t}\,\chi_{ij}) \rangle$, and the  renormalized exchange integral ($g^{s}J_{ij}$), where $g^{t}$ and $g^{J}$ are the
renormalization factors due to correlations:
\begin{equation}
g_{ij}^{t}=\sqrt{\frac{4x_{i}\,x_{j}(1-x_{i})(1-x_{j})}
{\left(1-x_{i}^{2}\right)\left(1-x_{j}^{2}\right)+8(1-x_{i}\,x_{j})|\chi_{ij}|^{2}+16|\chi_{ij}|^{4}}},
\end{equation}
\begin{equation}
g_{ij}^{J}=\frac{4(1-x_{i})(1-x_{j})}
{\left(1-x_{i}^{2}\right)\left(1-x_{j}^{2}\right)+8x_{i}\,x_{j}\,\beta_{ij}^{-}(2)+16\beta_{ij}^{\dagger}(4)},
\end{equation}
with $x_{i}\equiv 1-n_{i}$ and $\beta_{ij}^{\pm}(n)\equiv |\Delta_{ij}|^{n}\pm |\chi_{ij}|^{n}$.

Second, as noted earlier, Eq. (\ref{r12}) defines the correct renormalized mean-field approach in the sense that the
average fields $\Delta$ and $\chi$, derived from self-consistent equations coincide with those obtained variationally
from the appropriate Landau functional obtained from (\ref{r12}) \cite{b65} . In this manner, the Bogoliubov-Feynman
variational principle is obeyed.

The results displayed in Figs. 4b-d are drawn for different sets of model parameters (for details see \cite{b77}).
Irrespectively of the quantitative differences, few universal trends should be noted. First, there is a well defined
upper critical concentration contained in the regime of $x=1/4 \div 1/3$, in agreement with experimental results shows
in Fig. 4a. The presence of the upper critical concentration for disappearance of superconductivity has been interpreted
by us a signature of real space pairing \cite{b73}. Simply put, with the increasing hole doping the pairs get diluted to
the extent of destroying the condensed state.  Second, the trend of the data concerning the evolution with $x$ of the
gap parameter $\Delta_{\langle ij\rangle}$ in the antinodal ($k_{x},0$) direction is also reflected in trend of the
theoretical results. However, there is a problem with the $x$-dependence of the Fermi velocity in the nodal direction
($k_{x}=k_{y}$), as the experimental values are only weakly dependent on $x$. 
On the other hand, the theoretical data reflected the trend characteristic of a Fermi liquid, for which the diminishing
Fermi velocity with the decreasing $x$ is easy to understand. Therefore, the feature of the data collected in Fig. 4d
speaks in favor of a non-Fermi liquid and/or higher-order effects becoming eminent in the underdoped regime. We address
this discrepancy by discussing next the Gutzwiller-wave-function diagrammatic expansion (DE-GWF) in the next Section.

\subsection{Beyond the renormalized mean-field approach: systematic diagrammatic expansion for the
Gutzwiller-wave-function approach}

As said above, the Gutzwiller variational-method origin can be traced to the independent Hubbard-model inception
\cite{b47,b48}.  During the first 10 years (1965-1975), the Gutzwiller approximation (ansatz) was used frequently. After
that period, a systematic (iterative) solution for the full Gutzwiller wave-function has been achieved, first for a one
dimensional case \cite{b78,b79}. A generalization of this solution to the case of two spatial dimensions via systematic
diagrammatic expansion has been undertaken recently for both normal \cite{b80} and superconducting \cite{b81,b82}
states. In this brief overview we turn our attention to the most general features of the approach (for details see
\cite{b82}) and then concentrate on comparison with experiment.

The essence of developing a systematic expansion beyond any mean-field approximation is as follows. In any variational
procedure we would like to calculate the optimal grand-state energy $\langle \Psi|H|\Psi \rangle$. In the presence
approach
\begin{equation}
|\Psi\rangle \equiv |\Psi_{G}\rangle = \widehat{P}\, |\Psi_{0}\rangle =\prod_{i} \widehat{P}_{i}\, |\Psi_{0}\rangle
\end{equation}
where $\widehat{P}_{G}$ is the Gutzwiller projection operator, here taken in the form \cite{b83}
\begin{equation}
\widehat{P}_{i}= \sum_{\Gamma} \lambda_{i,\Gamma}\,|\Gamma\rangle_{i}\,_{i}\langle \Gamma |,
\end{equation}
with variational parameters $\lambda_{i,\Gamma}$ describing the occupation probabilities of four possible local (site)
configurations represented by the corresponding states, $\{ |\Gamma\rangle_{i} \} \equiv \{ |\emptyset\rangle_{i},
|\!\uparrow\rangle_{i}, |\!\downarrow\rangle_{i},|\!\uparrow\downarrow\rangle_{i}\}$. A choice of $\{ \lambda_{i\Gamma}
\}$ selected here provides the following form of the local projection operator
\begin{equation}
\widehat{P}_{i}^{2} \equiv 1+\tilde{x}\, \widehat{d}_{i}^{HF},
\end{equation}
where $\tilde{x}$ is a variational parameter and $\widehat{d}_{i}^{HF} =
(\widehat{n}_{i\uparrow}-\widehat{n}_{i\uparrow}^{0})(\widehat{n}_{i\downarrow}- \widehat{n}_{i\downarrow}^{0})$, with $
\langle n_{i\sigma}^{0} \rangle = \langle \Psi_{0}|\widehat{n}_{i\sigma}| \Psi_{0} \rangle$. With this form of the
projection operator one sees that the ground states energy can be expressed as
\begin{equation}
\langle\mathcal{H}\rangle = \frac{\left\langle \Psi_{0}\left| \widehat{P}\,\mathcal{H}\, \widehat{P} \right| \Psi_{0}
\right\rangle}{\left\langle \Psi_{0}\left|\,\widehat{P}^{2}\, \right|\Psi_{0} \right\rangle}.
\label{r18}
\end{equation}
So the evolution of the expectation value of $\mathcal{H}$ reduces to calculation of those for
$\widehat{P}\,\mathcal{H}\, \widehat{P}$  and $\widehat{P}^{2}$ in the $|\Psi_{0}\rangle$ which represents (and is
dependent on the problem and hand) single-particle wave function  $|\Psi_{0}\rangle$. This simplifies remarkably the
calculations, as the averages can be factorized into  products of simpler pair correlation functions using the Wick
theorem in real space. An efficient method of evaluating systematically the averages, including the pairing
correlations, has been overviewed in detail elsewhere in this issue \cite{b84}. Here we summarize only the main results.

The averages in (\ref{r18}) involve only the wave function $|\Psi_{0} \rangle$ representing an uncorrelated state.
Therefore, one can define an effective Hamiltonian $\widehat{\mathcal{H}}_{0}^{\rm eff}$ for which this state is an
eigenstate. A detailed analysis shows that both in the case of Hubbard \cite{b81} and t-J \cite{b82} model cases one can
define the effective single-particle Hamiltonian of the form:
\begin{eqnarray}
\widehat{\mathcal{H}}_0^{\rm eff} &=& \sum_{\pmb{i},\pmb{j}, \sigma}t^{\rm eff}_{\pmb{i},\pmb{j}}\,
\hat{c}_{\pmb{i},\sigma}^{\dagger}\,\hat{c}_{\pmb{j},\sigma}^{\phantom{\dagger}} + \sum_{\pmb{i},\pmb{j}} \left(
\Delta^{\rm eff}_{\pmb{i},\pmb{j}}\, \hat{c}_{\pmb{i},\uparrow}^{\dagger}\,\hat{c}_{\pmb{j},\downarrow}^{\dagger} +
\text{H.c.} \right), \label{r19}\\
t^{\rm eff}_{\pmb{i},\pmb{j}} &=& \frac{\partial \mathcal{F}\!\left(|\Psi_{0} \rangle,x\right) }{\partial
P_{\pmb{i},\pmb{j}}} \;, \qquad
\Delta^{\rm eff}_{\pmb{i},\pmb{j}} = \frac{\partial \mathcal{F}\!\left(|\Psi_{0} \rangle,x\right) }{\partial
S_{\pmb{i},\pmb{j}}} \;. \label{r20}
\end{eqnarray}
where $P_{\pmb{i},\pmb{j}} \equiv \langle \psi_{0}|\hat{c}_{i\sigma}^{\dagger}\, \hat{c}_{j\sigma}| \psi_{0} \rangle -
\delta_{ij}\langle \Psi_{0}|\hat{n}_{i\sigma}|\Psi_{0} \rangle$, $S_{ij}\equiv  \langle
\psi_{0}|\hat{c}_{i,\uparrow}^{\dagger}\, \hat{c}_{j,\downarrow}^{\dagger}| \psi_{0} \rangle $ and $ \mathcal{F}\equiv
\langle \psi|\mathcal{H}|\psi  \rangle_{G}-2\mu_{G}\,n_{G} $ with $n_{\sigma}\equiv\langle
\Psi_G|\hat{n}_{i\sigma}|\Psi_{G} \rangle$ is the generalized grand-canonical potential in the GWE state. One see that
(\ref{r19}) is the effective single-particle Hamiltonian with BCS-type pairing in the real space-language. This equation
contains a number of parameters to be determined self-consistently: $ P_{i,j}, \Delta_{i,j}, \mu$, and $n_{G}$ (for a
detailed discussion see Refs. \cite{b81,b82}). In Fig. 5a-d we have assembled the principal results of the approach.

Explicitly, Fig 5a illustrates the two features of the solution. First, superconducting state appears only at
sufficiently high $U$ ($U\equiv U/|t|\gtrsim 3$). This shows explicitly that in obtaining a stable superconducting
solution the electron correlations must be taken into account, as there is no such solution in the Hartree-Fock
approximation. Also, the BCS/non-BCS boundary specifies the shaded regime in the lower right-hand corner, where the
kinetic energy ($\Delta E_{kin}$) is lowered in the superconducting phase. The line within the shaded regime marks the
situation in which the potential energy gain forming SC phase is zero, a clearly non-BCS feature. In Fig. 5b we
displayed evolution of the doping dependence of the correlated gap with the increasing $U$. We see that for sufficient
high $U$-value, the system transforms into the Mott insulator at $\delta=0$, as it should be. The lower panel on that
figure demonstrates the convergence of the results for $\Delta_{G}$ with the ascending order of the DE-
GWF expansion. The results for $k=4$ and $k=5$ practically coincide, a rewarding feature in view of the intricacy of the
expansion \cite{b84}. In Fig. 5c we draw the $\pmb{k}$-dependence of the gap and, in particular, demonstrate the
deviation from a pure $d$-wave solution for $\Delta_{\pmb{k}}$ away from the optimal doping. This type of behavior is
observed experimentally \cite{b85}, though a quantitative comparison with experiment would involve also determination of
the doping dependence of the pseudogap. Finally, in Fig. 5d the Fermi velocity is plotted against $\delta$, with the
same experimental points, as in Fig. 4d. We see that the theory reflects now to much better extent the trend of the
experimental data in the last Figure. Indeed, DE-GWE approach represents more advanced approach than SGA. It should be
underlined though that while the plots Fig. 4 illustrate the SGA results for t-J model, those in Fig. 5 are obtained for
the Hubbard model. What is important, SGA does not produce a stable 
superconducting state within the Hubbard model.

Recently, we have also reanalyzed the t-J model within the DE-GWF approach (for details see \cite{b82}) and the results
are of similar character as those for the Hubbard model. Explicitly, in Figs. 6a-d we summarize those results and
compare them to those obtained in Variational Monte-Carlo (VMC) method (cf. Fig. 6a). Figs. 6b-d can be directly
compared to those of Figs. 4b-d. Here SGA $\equiv$ GCGA (\emph{grand canonical Gutzwiller approximation}). One can see
that the $v_{F}(\delta)$ dependence is best reproduced within the Hubbard model (cf. Fig. 5d).

On the basis of the results depicted in Figs. 3-5 we see that the strong correlations are necessary to reproduce an
overall behavior. Note that the lowest points in Fig. 4d are those obtained from DMFT approach \cite{b86} for the same
values of the parameters. Additionally, the results presented here are of the same quality as those obtained from
Variational Monte Carlo method \cite{b82}. All these features prove that the DE-GWF (starting from SGA as the lowest
order) is a reliable method for treating the high-temperature superconductivity, at least its overall features.

\section{From Anderson to Anderson-Kondo lattice}

\subsection{Anderson-Kondo lattice}

At the end we overview briefly the evolution of the Anderson-lattice model from description of quasiparticle states in
heavy-fermion systems and their magnetism to the discussion of paired superconducting states. To model the heavy fermion
systems, such as the compounds with cerium (with approximate $4f^{1}$ configuration of Ce$^{3+\delta}$ ions), we start
from the two-orbital Anderson lattice model, which is the real-space representation has the following form:
\begin{eqnarray}
\mathcal{H} =& &\sum_{mn\sigma}\,t_{mn}\, \hat{c}^{\dagger}_{m\sigma}\, \hat{c}_{n\sigma} + \epsilon_f \sum_{i\sigma}
\hat{N}_{i\sigma} + U \sum_{i\sigma} \hat{N}_{i\uparrow}\,
\hat{N}_{i\downarrow} \nonumber\\
&&+ \sum_{im\sigma} \left( V_{im}\, \hat{f}^{\dagger}_{i\sigma}\, \hat{c}_{m\sigma} + \text{H.c.} \right) - \mu \left(
\sum_{i\sigma} \hat{N}_{i\sigma} + \sum_{m\sigma} \hat{n}_{m\sigma}\right),
\label{r21}
\end{eqnarray}
where $\hat{N}_{i\sigma}\equiv \hat{f}^{\dagger}_{i\sigma}\,\hat{f}_{i\sigma}$ is the number of $f$ electrons on site
$i$ with spin $\sigma$, and $\hat{n}_{m\sigma}\equiv \hat{c}^{\dagger}_{m\sigma}\,\hat{c}_{m\sigma}$ is the
corresponding number of the conduction ($c$) electrons. The meaning of the consecutive terms is as follows: the first
term expresses band (hoping) energy of $c$-electrons; the second, the starting atomic energy of $f$ electrons (with
their level position $\epsilon_{f}$); the third the $f$-$f$ intraatomic (Hubbard) interaction; the fourth the $f$-$c$
hybridization (with amplitude $V_{im}$); and the last, the corresponding chemical potential part, as we work in the
grand-canonical formalism.

In the interesting us regime of strong correlations the parameter $U$ represents the highest energy scale in the system.
In the regime, when the emerging quasi-particle states involve itineracy of $f$ electrons, neither the RKKY $f$-$f$
interaction \cite{b2} nor the Schrieffer-Wolff transformation \cite{b19} of the Anderson-lattice model to the effective
Kondo-lattice model, are applicable. Instead, at best one can invoke the concept of the \emph{Anderson-Kondo lattice},
which we explain in detail and apply subsequently to the description of paired states.

It is important to note that while the ratio $|V|/(U+\epsilon_{f})\ll 1$, the relative hybridization strength
$V/\epsilon_{f}$  cannot be regarded as small and therefore transformed out, as would be the case in the situation with
the Schrieffer-Wolff transformation. This circumstance leaves  us always with a residual hybridization even when we
transform canonically the starting Hamiltonian (\ref{r21}) into that containing the Kondo-type interaction in an
explicit form. Leaving the details of such transformation aside \cite{b87,b23,b22}, the effective Hamiltonian under
these assumptions has the form
\begin{eqnarray}
\mathcal{H} = & & \mathcal{\hat{P}} \left\{
\sum_{mn\sigma} \left(t_{mn}\, \hat{c}^{\dagger}_{m\sigma}\, \hat{c}_{n\sigma} -\sum_{i} \frac{V_{im}^{*}\, V_{in}}{U +
\epsilon_f}\, \hat{\nu}_{i\bar{\sigma}}\,\hat{c}^{\dagger}_{m\sigma}\, \hat{c}_{n\sigma}  \right)
 \right\} \mathcal{\hat{P}}  \nonumber\\
+&&
 \mathcal{\hat{P}} \left\{\sum_{imn \sigma} \frac{ V_{im}^{*}\, V_{in} }{U + \epsilon_f}\,\hat{S}_{i}^{\sigma}
\hat{c}^{\dagger}_{m\bar{\sigma}}\, \hat{c}_{n\sigma} + \sum_{i\sigma}{\epsilon_f }\, \hat{\nu}_{i\sigma}  \right\}
 \mathcal{\hat{P}}\nonumber\\
+&&
 \mathcal{\hat{P}} \left\{ \sum_{im\sigma} \left(1-\hat{N}_{i\bar{\sigma}}\right) \left(V_{im}\,
\hat{f}^{\dagger}_{i\sigma}\, \hat{c}_{m\sigma} +  \text{H.c.}\right) \right\}
 \mathcal{\hat{P}}\nonumber\\
+&&
 \mathcal{\hat{P}} \left\{
\sum_{im\sigma} \frac{2 |V_{im}|^2
  }{U+\epsilon_f} \left(\pmb{\hat{S}}_i\cdot \pmb{\hat{s}}_m - \frac{\hat{\nu}_{i}\,\hat{n}^c_{m}}{4}  \right)  \right\}
 \mathcal{\hat{P}}\nonumber\\
+&&
 \mathcal{\hat{P}} \left\{
\sum_{ij} J_{ij} \left(\pmb{\hat{S}}_i\cdot \pmb{\hat{S}}_j - \frac{\hat{\nu}_{i}\, \hat{\nu}_{j}}{4} \right)  \right\}
\mathcal{\hat{P}}.
\label{r22}
\end{eqnarray}

Note that in the hybridization term we have projected out only the processes with the double $f$-level occupancies (cf.
third line), whereas both the effective Kondo ($f$-$c$) (the fourth line) and $f$-$f$ exchange (superexchange)
interaction (the last line) appear in an explicit form as higher-order processes. One should underline that the form
(\ref{r22}) should contain the same physics as (\ref{r21}) in the limit we call the Anderson-Kondo limit. However, as we
see shortly, it provides physically appealing solutions already within a relatively simply approximation scheme.

Before discussing the results, we introduce explicitly the real-space pairing operators in the following manner
\begin{equation}
\left\{\begin{aligned}
\hat{b}^\dagger_{im} &\equiv  \frac{1}{\sqrt{2}} \left(\tilde{f}^\dagger_{i\uparrow}\, \hat{c}^\dagger_{m\downarrow} -
\tilde{f}^\dagger_{i\downarrow}\, \hat{c}^\dagger_{m\uparrow}\right)\equiv (\hat{b}_{im})^{\dagger}, \\
\hat{B}_{ij}^{\dagger} &\equiv  \frac{1}{\sqrt{2}} \left( \tilde{f}_{i\uparrow}^{\dagger}\,
\tilde{f}_{j\downarrow}^{\dagger} -\tilde{f}_{i\downarrow}^{\dagger}\, \tilde{f}_{j\uparrow}^{\dagger}\right) =
\left(\hat{B}_{ij}^\dagger\right)^\dagger,
\end{aligned}
 \right.
\end{equation}
where $\tilde{f}_{j\sigma}\equiv \hat{f}_{i\sigma}(1-\hat{N}_{i\bar{\sigma}})$ and $\hat{\nu}_{j\sigma}\equiv
\hat{N}_{i\sigma}(1-\hat{N}_{i\bar{\sigma}})$. In effect, the Hamiltonian (\ref{r22}) can be rewritten in a closed form
\begin{eqnarray}
\mathcal{H} = &&\sum_{mn\sigma} t_{mn}\, \hat{c}^\dagger_{m\sigma}\, \hat{c}_{n\sigma} +
\epsilon_f\sum_{i\sigma}\hat{\nu}_{i\sigma} + \sum_{im\sigma}\left( V_{im}\,
\tilde{f}^\dagger_{i\sigma}\,\hat{c}_{m\sigma}+\text{H.c.}\right)\nonumber\\
&&- \sum_{imn} \frac{2 V^*_{im}\,V_{in}}{U+\epsilon_f}\, \hat{b}^\dagger_{im}\,\hat{b}_{in} - \sum_{ij} J_{ij}\,
\hat{B}^\dagger_{ij}\, \hat{B}_{ij}.
\label{r24}
\end{eqnarray}
The first term in the second line represents the so-called \emph{hybrid real-space pairing} and express the Kondo-type
spin-singlet correlations, whereas the second the corresponding $f$-$f$ pairing of the type considered already in the
context of t-J model.
The real-space pairing parts diminishe the system energy in the spin-singlet paired state. The pairing
$\hat{b}^{\dagger}\hat{b}$ introduces the hybrid ($f$-$c$) local-pair contribution and $\hat{B}^{\dagger}\hat{B}$
expresses the corresponding local $f$-$f$ binding. Those two types can compete and lead to a frustration effects in the
paired state.

\subsection{Magnetic and paired states: phase diagram}

The Hamiltonian has been solved within the SGA scheme \cite{b22,b23} and some of the results (and the whole method) are
over-viewed briefly in Fig. 7a-d. Namely,  in Fig 7a we visualize the basis on which the transformation from (\ref{r21})
to (\ref{r22}) has been carried out. The high-energy processes lead to the $f$-$c$ and $f$-$f$ exchange interactions;
the low-energy correspondants represent the residual hybridization. Fig. 7b illustrates that, strictly speaking, three
physically distinct regimes in large-$U$ limit should be singled out, as specified, with the increasing ratio
$V/\epsilon_{f}$. In Fig. 7c we show exemplary result in the Kondo-insulator regime (for $n_{e}=2$), where the quantum
critical point (QCP) appears between the antiferromagnetic Kondo insulator (AKI) and the nonmagnetic Kondo insulator
(PKI), the latter with totally compensated magnetic moments of $f$ ($m_{f}$) and $c$ ($m_{c}$) electrons. One should
emphasize that such a completely compensated state is due to the two factors: the 
antiferromagnetic Kondo interaction from one side and the autocompensation of itinerant character of $f$ electrons
combined with the antiferromagnetic kinetic exchange between them from the other. The inset shows the diminishing
$f$-level occupancy $n_{f}\equiv \sum_{\sigma} \langle \hat{\nu}_{i\sigma} \rangle$ with the increasing strength of the
(intraatomic in this case) hybridization.

Finally, in Fig. 7d we provide the overall phase diagram, this time for the case with hybridization between the
nearest-neighboring pairs $\langle  i,m\rangle$. The superconducting solution is of the $d$-wave type, both for the
hybrid and $f$-$f$ Cooper pairs. The sequence of the phases with the increasing nearest-neighbor hybridization amplitude
$|V|=|V_{\langle  i,m\rangle}|$ is as follows. For small hybridization the strong (SFM) and the weak (WFM) ferromagnetic
phases appear in this regime of practically localized $f$ electrons ($n_{f}>0.9$). A purely antiferromagnetic (AF) phase
is sandwiched in between the mixed antiferromagnetic-superconducting (AF+SC) phases. For large enough $|V|$, a pure
superconducting phase emerges in the fluctuating-valence regime, $n_{f}\simeq 0.8$. The sequence of the phases near QCP
(the black solid dots in Figs. 7c and d) reflects (inverted $"V"$) quantum-critical behavior appearing in the
quasi-two-dimensional heavy-fermion system \cite{b88} (the results are calculated for 
the case of square lattice), dividing the AF and SC phases, and with AF+SC phase inside.

Concluding this Section, the Anderson-Kondo lattice model leads to a number of magnetic and superconducting phases, both
induced by the magnetic interactions and the interelectronic correlations combined. A separate question concerns the
DE-GWF generalization for the Anderson-Kondo lattice \cite{b89}, but this topic will not be touched upon here.

\section{Outlook}

In this brief overview we have put an emphasis on the connection between magnetic and superconducting states, both
treated on the same footing and within a single model of correlated fermions. In other words, no extra fermion-boson
interaction is necessary to introduce both magnetism and/or unconventional superconductivity. The question is to what
extent the above models conceived more than 50 years ago convey still novel and relevant physics of many-particle
systems. One may say that what wee still need is their systematic analysis for the case of periodic lattices,
particularly for orbitally degenerate cases of $d$ and $f$ states. This goal should lead to a formulation of
universality classes for continuous quantum phase transitions, definition of the upper and the lower critical dimensions
for them, as well as a unified view of the spin correlations in strongly correlated fermionic systems and their relation
to the pairing in high-temperature and heavy-fermion superconductors. The quantum  Monte-Carlo 
methods in this respect provide a crucial testing ground for various approximate analytical/numerical solutions in the
situation of small systems, but the final goal is to have the solution for extended (infinite) systems. Whether in
achieving this goal we should have first a renormalized single-particle approach along the lines discussed here, remains
still to be seen. For example, apart from a single result shown in Fig. 7c, no quantum criticality in the low-$U$ limit
has has been touched upon here \cite{b90, b91,b92}. The consideration of quantum criticality leads to new physics
(\emph{non-Fermi liquid behavior}), for both models. Also, the particular questions of the pseudogap appearance
\cite{b93} in high-$T_{c}$ systems and, e.g., the hidden-order existence in URu$_{2}$Si$_{2}$ \cite{b94}, are not
tackled here. Nonetheless, I firmly believe that the  considerations touched upon here provide a first if not
substantial step in understanding theoretically the superconductivity in the strong-correlation 
limit for the narrow-band fermions.

\section*{Acknowledgment}

I am very grateful to my former and present Ph.D. students: Jakub J\c{e}drak, Jan Kaczmarczyk, Olga Howczak, Marcin
Abram, Marcin Wysoki\'{n}ski, and Ewa K\c{a}dzielawa-Major, for discussions and making available some of the results of
our joint or still unpublished works. I thank also to Danuta Goc-Jag{\l}o for her technical help. This work was
supported by the Foundation for Polish Science (FNP) through Grant TEAM, as well as by the National Science Centre (NCN)
through Grant MAESTRO, No. DEC-2012/04/A/ST3/00342. I thank to Peter Riseborough for his help in editing this and other
articles in this special issue.

\begin{figure}
\begin{center}
\includegraphics[width=7cm]{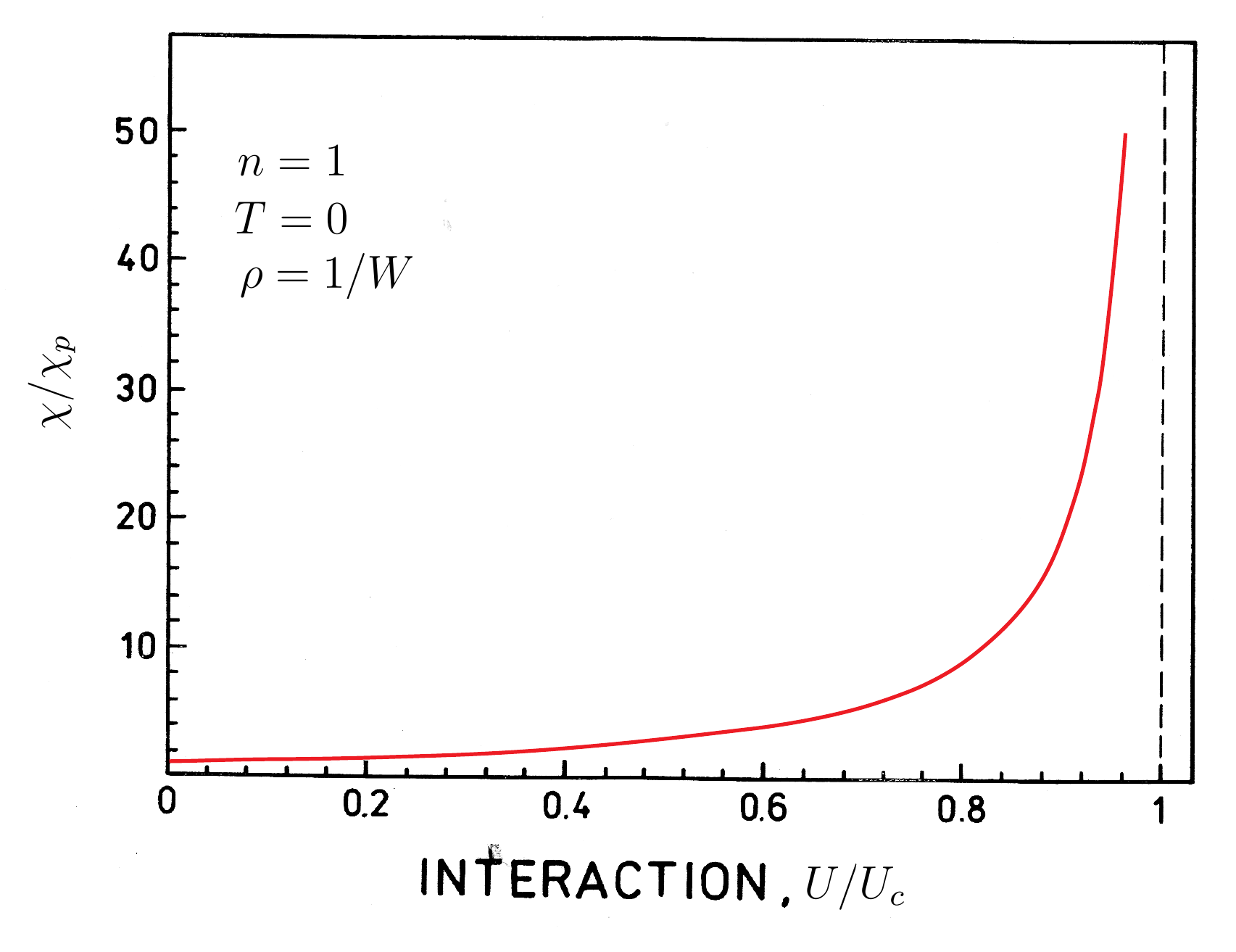}
\caption{(Color online) Relative static paramagnetic zero-field susceptibility per site ($\chi/\chi_{p}$) as a function
of the relative intraatomic interaction strength
$U/U_{c}$ for a half-filled narrow band. The divergence marks the Mott-Hubbard localization, called also the
Brinkman-Rice instability point.}
\end{center}
\end{figure}

\begin{figure}
\begin{center}
\includegraphics[width=7cm]{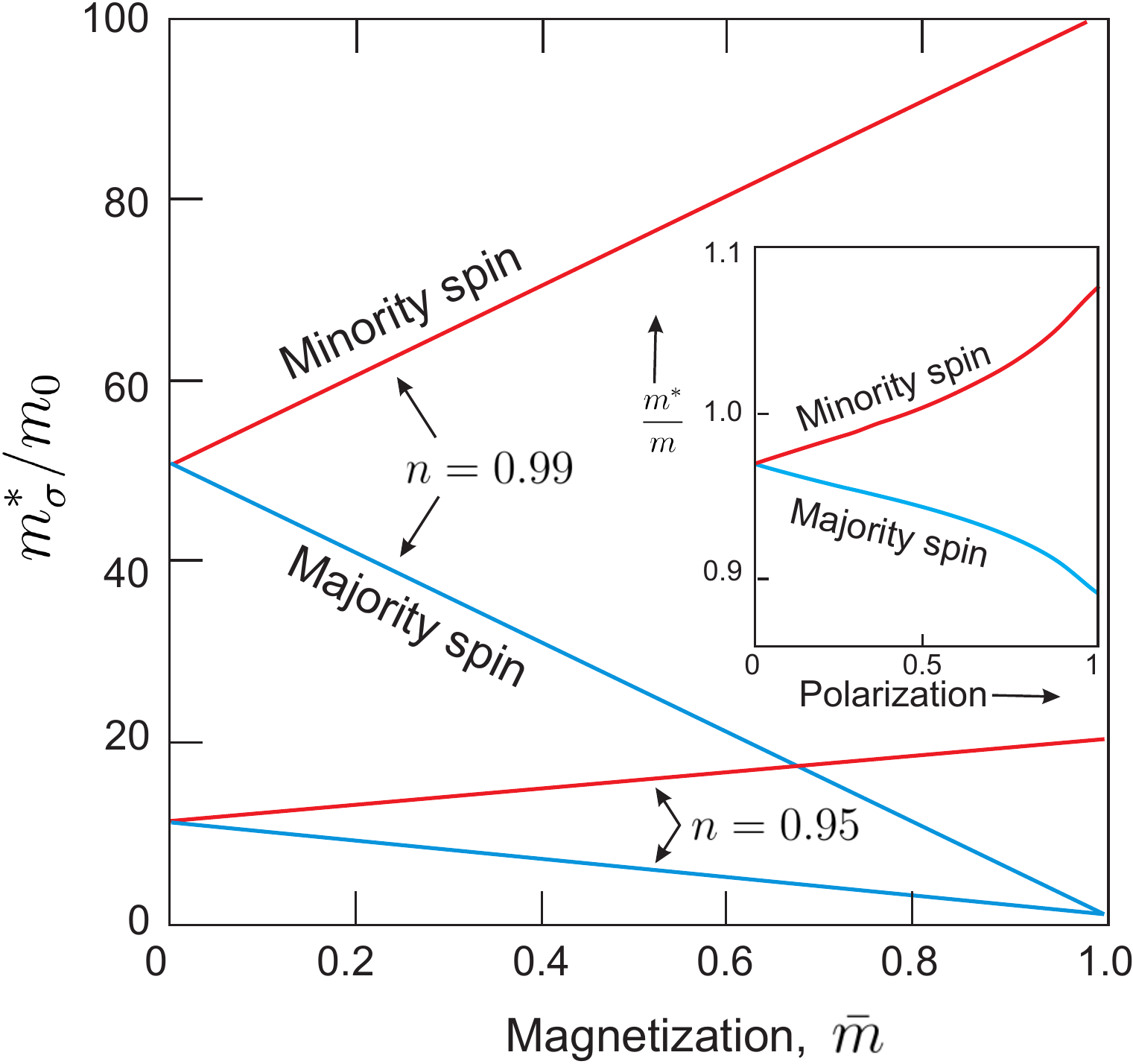}
\caption{(Color online) The enhancements of spin-split masses $m^{*}_{\sigma}/m_{0}$ for $n<1$ as a function of relative
magnetic moment $\bar{m}=m/n$ for the two band fillings specified. The inset shows the corresponding dependence for the
electron gas (cf. Spa{\l}ek and Gopalan, 1990).  Note that the majority-spin carriers acquire the bare band mass as the
magnetic saturation is reached, whereas those in the spin minority band become extremely heavy and disappear as the  
state is approached}
\end{center}
\end{figure}

\begin{figure}
\begin{center}
\includegraphics[width=12cm]{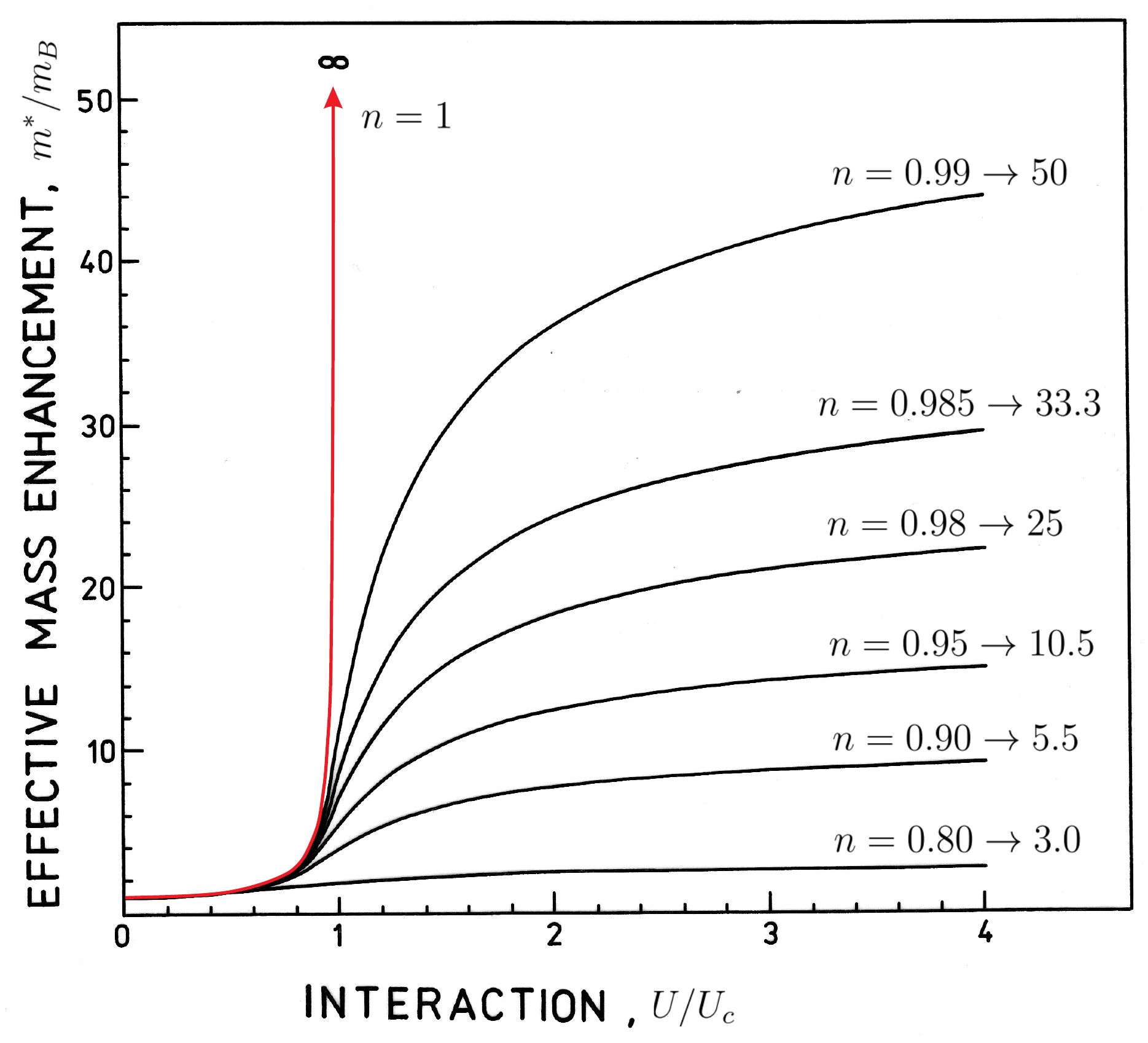}
\caption{(Color online) Effective mass enhancement versus $U/U_{c}$. On the right: the asymptotic values of
$m^{*}/m_{B}$ for $U/U_{c}\rightarrow \infty$ and for selected values of band fillings $n$. All the curves are drawn for
a constant form of the density of states ($\rho^{0}(\epsilon)=1/W$, where $W$ is the bare bandwidth.}
\end{center}
\end{figure}

\begin{figure}
\begin{center}
\includegraphics[width=12cm]{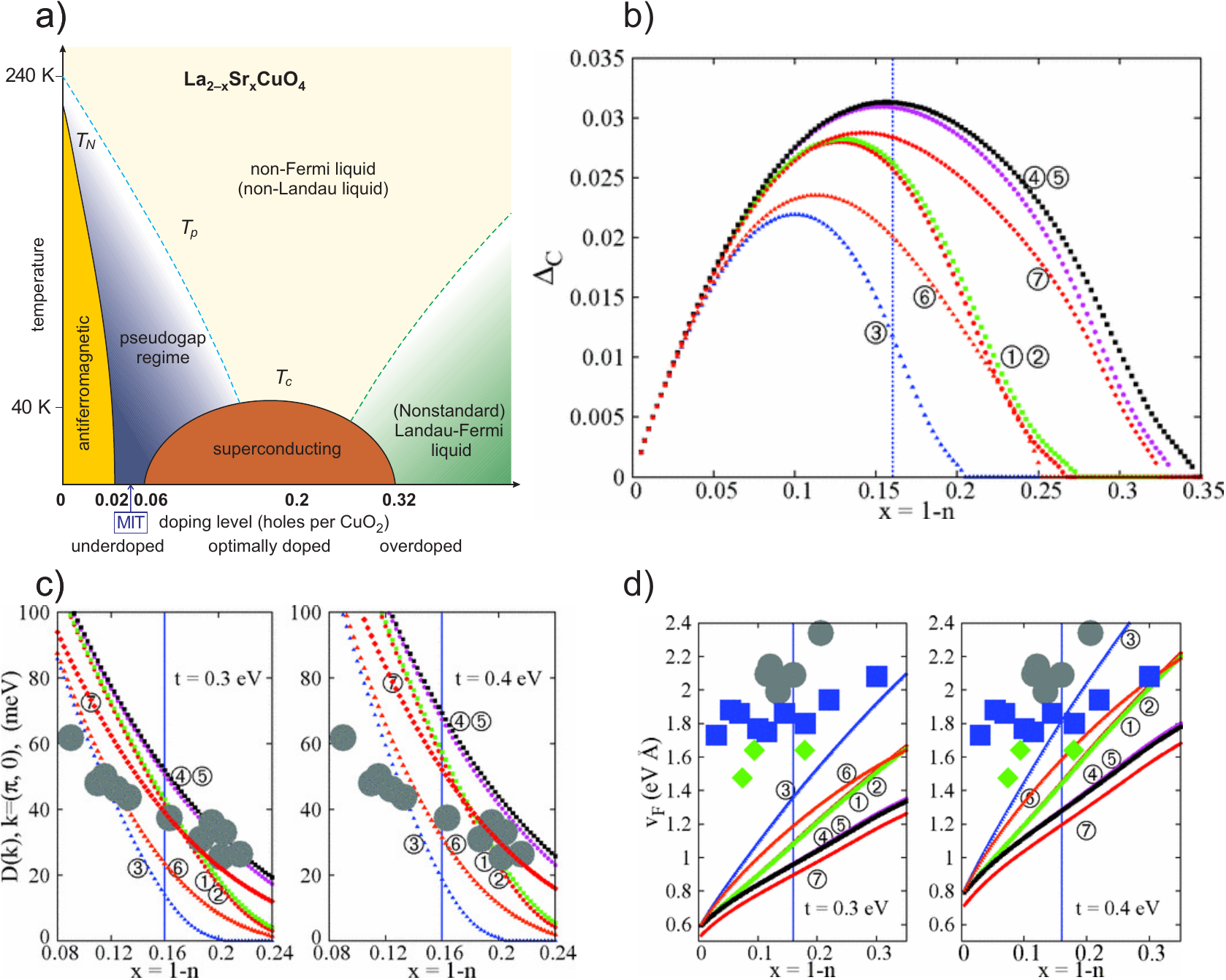}
\caption{(Color online) a) representative phase diagram for cuprate high $T_{C}$ superconductors: $T_{p}$ means
pseudogap characteristic temperature and $T_{C}$ is the critical temperature for superconducting phase transition; b)
theoretical phase diagram specifying only the superconductivity to normal metal transition; c) the gap amplitude versus
doping $x$  in the antinodal ($k_{x},0$) direction for the two values of the nearest neighbor parameter $t\equiv |t|$
specified; d) Fermi velocity versus $x$ in the nodal direction $k_{x}=k_{y}$; note the systematic derivation in the
underdoped regime (the optimal doping is marked on Figs. b-d as a vertical line).}
\end{center}
\end{figure}

\begin{figure}
\begin{center}
\includegraphics[width=12cm]{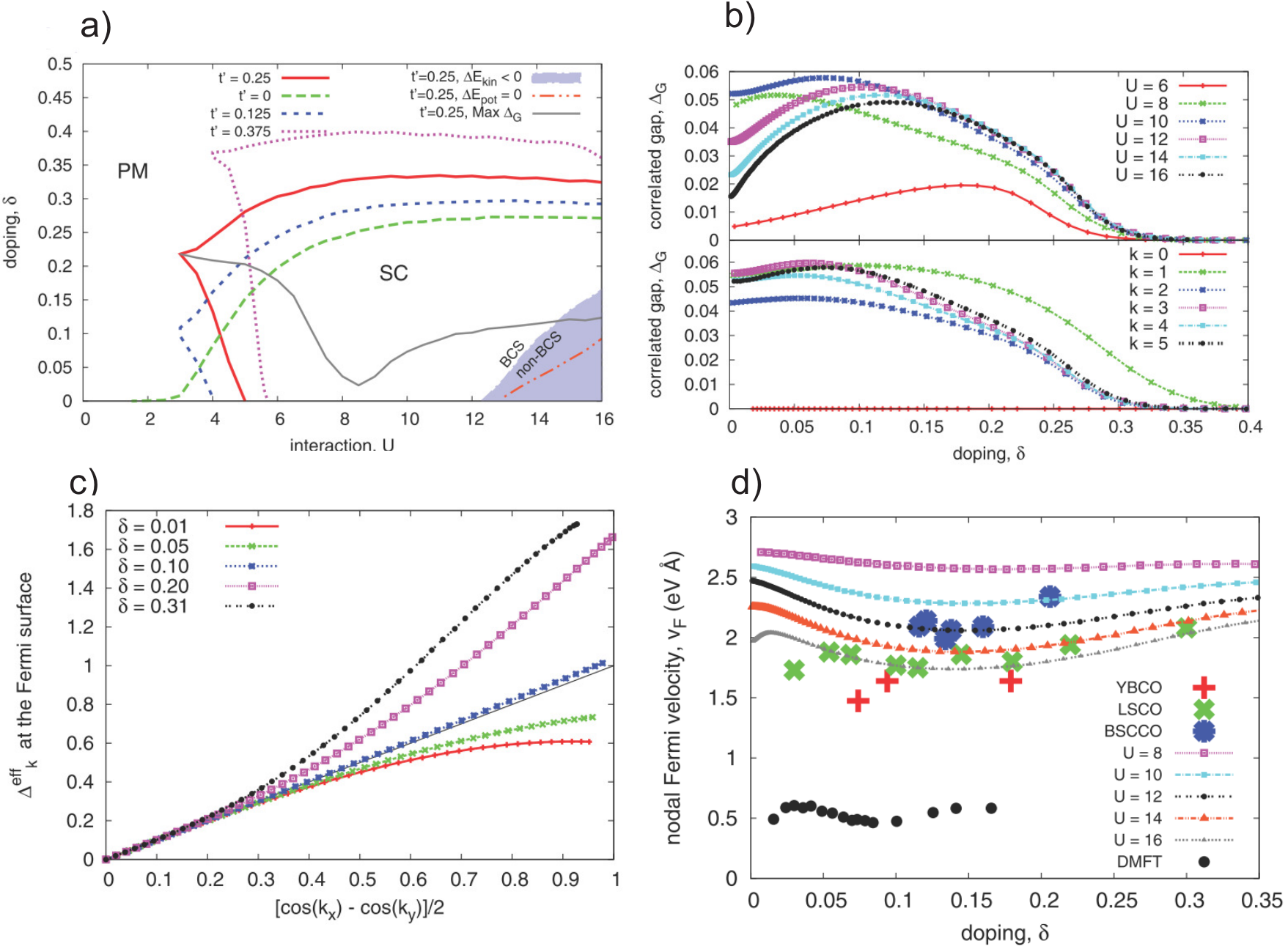}
\end{center}
\caption{(Color online) a) Stability regime on plane interaction $U-$ doping$\delta$ for specified values of the
parameters (in the lower right corner the BCS vs. non-BCS regimes, see main text); b) superconducting gap magnitude vs.
$\delta$ for different $U$ (the lower panel: the convergence with the order $k$ of thr DE-GWF expansion) c) correction
to the poor d-wave gap due to longer-range pairing components (the black solid line marks a purse d-wave character); d)
Fermi velocity vs. $\delta$: the upper set of points are the are the some experimental data as in Fig. 4d, whereas the
lower solid points represent DMFT results. For details see Ref. \cite{b81}}
\end{figure}

\begin{figure}
\begin{center}
\includegraphics[width=12cm]{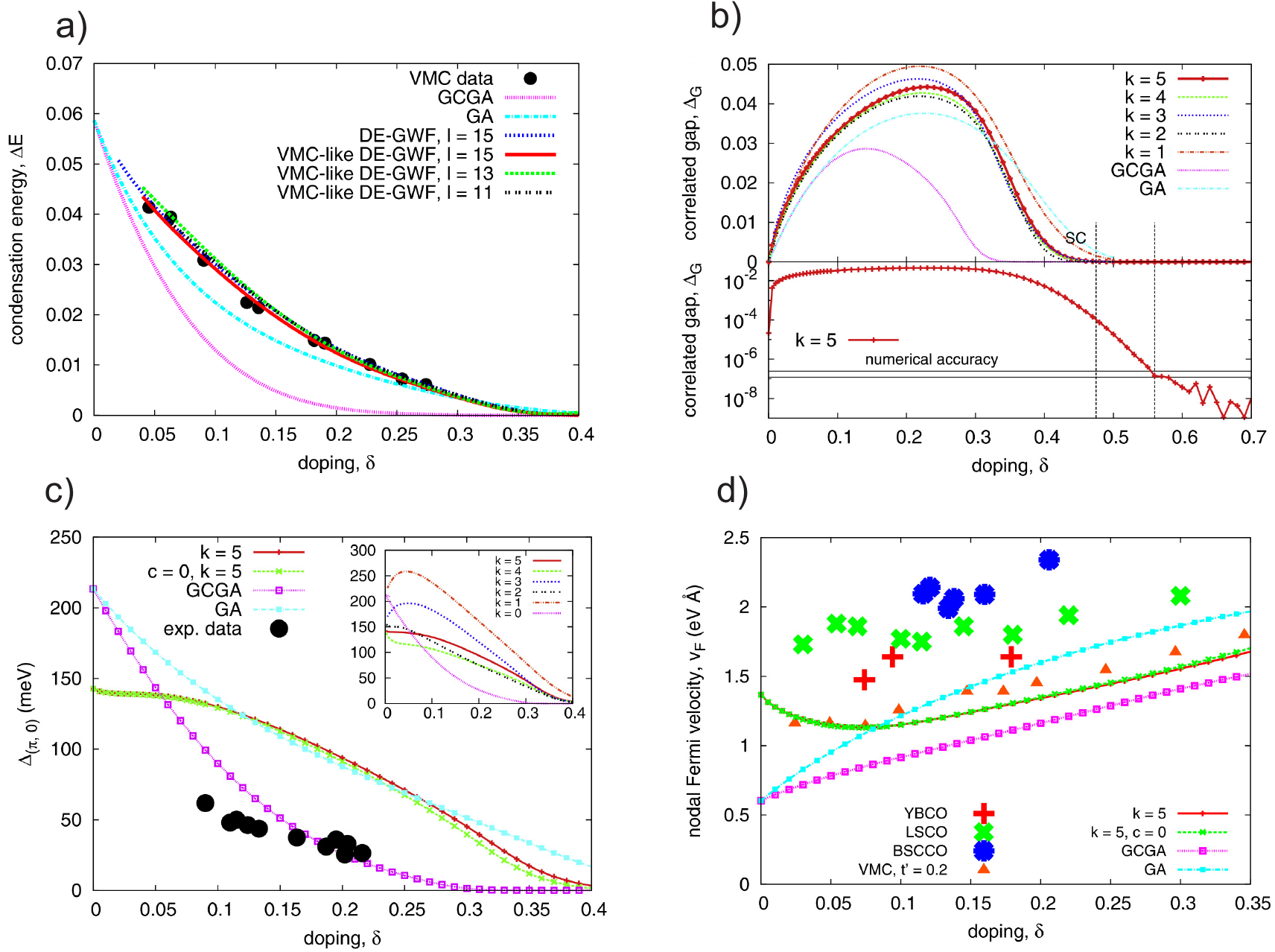}
\end{center}
\caption{(Color online) Characteristics of the paired superconducting state coming from the t-J model and obtained in
DE-GWF method \cite{b82} as a function of hole doping $x=\delta$: 
a) condensation energy compared with variational Monte-Carlo (VMC) and  VMC-like results, the latter contain the
diagrams with a $ l=15,13$, and 11 lines; the lowest curve represents SGA results, whereas the black points those from
VMC;
b) the correlated gap in the different order of the DE-GWF expansion (the lowest one represents SGA results); the lower
part shows the numerical accuracy of $\Delta_{G}$ evaluation in both $\delta\rightarrow 0$ and $\delta>0.4$; 
c) the gap amplitude in the antinodal direction in different orders (the lowest curve represents SGA results and is
close to the experimental results;
d) universal Fermi velocity in the nodal direction and its comparison to experiment.}
\end{figure}

\begin{figure}
\begin{center}
\includegraphics[width=12cm]{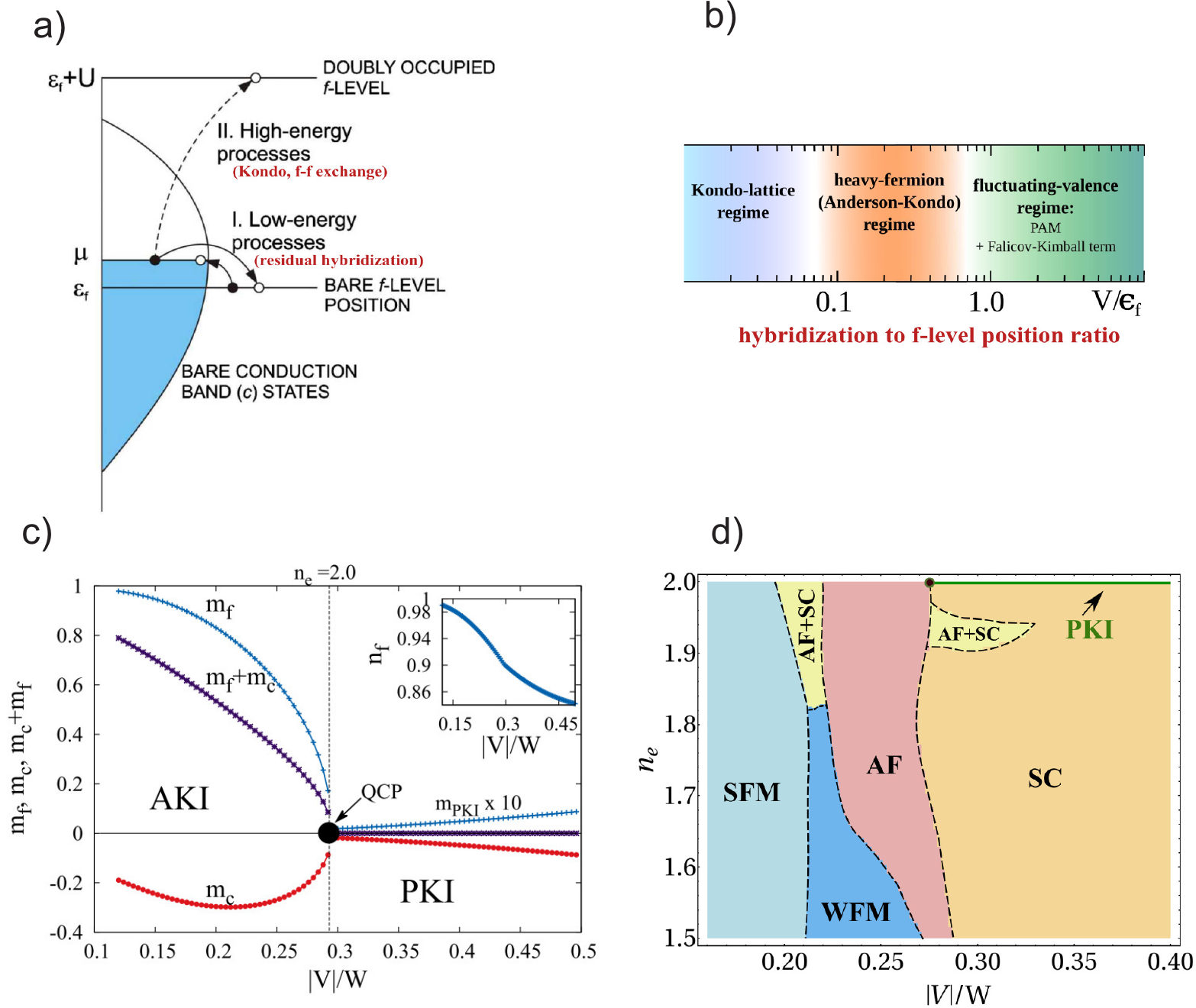}
\end{center}
\caption{(Color online) a) Schematic representation of hybridization $c$-$f$ process division into the low- and
high-energy processes; b) Various physical regimes for Anderson-lattice model with the increasing $V/\epsilon_{f}$
ratio; c) The compensation of the magnetic moment ($m_{f}$) by that of conduction electrons ($m_{c}$) in the state with
itinerant $f$-quasiparticles; d) overall phase diagram on the plane number of electrons per pair of orbits ($n_{e}$)
-hybridization strength ($|V|$). For details see main text and Ref. \cite{b23}
}
\end{figure}

\end{document}